\documentclass[10pt,letterpaper,authoryear]{elsarticle}
\usepackage{graphicx}% Include figure files
\usepackage{rotating}
\usepackage{nameref}
\usepackage{bm}% bold math
\usepackage{amssymb}
\usepackage{hyperref}
\usepackage{cite}
\usepackage{amsmath}
\usepackage{natbib}
\usepackage{dcolumn,multicol}% Align table columns on decimal point
\begin{document}
%\begin{frontmatter}
%
%\title{Changes in sinking of plankton--like particle: comparison between observations and numerical model} 
%
%\author{E.Y.Shchekinova$^{1,2}$,Christina Geb{\"u}hr$^{1}$, Maarten Boersma$^{1}$, Karen H. Wiltshire$^{1}$}
%
%\ead{elena.shchekinova@gmail.com}
%\address{$^1$ Alfred--Wegener--Institut Helmholtz--Zentrum f{\"u}r Polar-- und Meeresforschung, Biologische Anstalt Helgoland, Kurpromenade 201, D-27498 Helgoland, Germany\\
%$^2$ Stuttgart Research Center System Biology, University of Stuttgart,
%Nobelstra{\ss}e 15, 70569 Stuttgart, Germany\\
%$^3$ Institute of Geosciences, Goethe--University Frankfurt, 60438 Frankfurt am Main, Germany
%}
%\date{\today}% It is always \today, today,
%             %  but any date may be explicitly specified
\vspace*{0.35in}

\begin{flushleft}
{\Large
\textbf\newline{Modeling of small sea floaters  in the central Mediterranean Sea: seasonality of at--sea distributions}
}
\newline
% authors go here:
\\
E.Y.Shchekinova\textsuperscript{1,2},
 Y. Kumkar\textsuperscript{3}
\\
\bigskip
\bf{1}Stuttgart Research Center for Systems Biology, University of Stuttgart, Germany  
\\
\bf{2} International Institute for Applied Systems Analysis (IIASA), Austria
\\
\bf{3} Norwegian Institute of Bioeconomy Research, Norway\\
\bigskip
* shchekinova.elena@gmail.com

\end{flushleft}

%\authorrunning{Short form of author list} % if too long for running head

%\institute{S. Galliani, E.Y. Shchekinova, T. Joshi, M. Reuss  \at
%              Stuttgart Research Center System Biology, University of Stuttgart,
%              Nobelstra\ss e 15, 70569 Stuttgart, Germany\\
%              Tel.: +123-45-678910\\
%              Fax: +123-45-678910\\
%              \email{shchekinova.elena@gmail.com}           %  \\
%%             \emph{Present address:} of F. Author  %  if needed
%          % \and
%          % E. Y. Shchekinova \at              
%}
%\date{Received: date / Accepted: date}
% The correct dates will be entered by the editor

%\maketitle

\section{Abstract}
Floating marine debris represent a threat to marine  and coastal ecology. Since the Mediterranean basin is one of the highly impacted regions, both by the coastal pollution as well as from sea traffic, the potential harm of a floating pollution on the marine ecology could be overwhelming in this area.  Our study area covers the central Mediterranean crossing that connects the western and eastern Mediterranean and is one of the areas impacted by a high intensity of sea traffic. To identify regions in the central Mediterranean that could be more exposed  by high concentration of floating marine pollutants we use Leeway model for lower windage small--size particles. We perform numerical simulation of a large ensemble of Lagrangian particles that approximate at---sea debris. The particles are forced by  high--resolution sea kinematics from the Copernicus Marine Environment Monitoring Service (CMEMS) and $10$ m atmospheric wind from the European Centre for Medium--Range Weather Forecasts (ECMWF) for two reference periods in summer and winter of 2013--2016. We identify the regions with a high accumulation of particles  in terms of particle surface densities per unit area. Although seasonal and annual variability of ocean current and atmospheric wind is an important factor that influences accumulation regimes across the central Mediterranean, we found that the border of the Libyan shelf  harbors larger percentage of particles after $30$ days of simulation.
%\keywords{Lagrangian transport \and Marine Pollution\and Leeway drift \and Mediterranean surface currents}
 \section{Introduction}
In the enclosed basin like the Mediterranean sea pollution can significantly impact the ecosystem health. Accumulation of plastic material can alternate microbial environment and harbor toxic and potentially hostile bacteria. Even short--term retention of waste in certain areas could affect the underwater climate and ecosystem diversity. Due to a higher concentration of floating waste the light climate and, consequently, underwater community can be affected, especially this concerns light--dependent plankton species. 

The existence of hot--spots of a floating waste accumulation was already demonstrated in many studies \citep{Lebreton2012,Mansui2014,Zambianchi2017,Liubartseva2018}. A study by Lebreton et al. (2012) based on a long--term simulation of Largangian particles coupled to the global ocean circulation model confirmed the existence of the large--scale aggregation zones. For the semi--enclosed basin such as the Mediterranean Sea floating particles were shown to accumulate along the north African coast in the central Mediterranean sea (CMED) and in eastern Mediterranean \citep{Lebreton2012}. In the models of Lagrangian transport of surface drifters with higher resolution data \citep{Mansui2014,Zambianchi2017} an evidence of a large spatial heterogeneity in the distribution of surface marine drifters across the CMED and in the eastern Mediterranean was found. Numerical study by  Mansui et al. (2014) based on multi--annual simulations of surface passive debris distributed across the entire Mediterranean identified several retention areas: the Gulf of Sirte and entire coastal strip from Tunisia to Syria. A recent study of tracking of Lagrangian particles showed that long term accumulation patterns  in the Mediterranean basin could not be found due to an average particle life time of $7-80$ days and an overall dissipative behaviour with respect to floaters at sea surface \citep{Liubartseva2018}.

\begin{figure}
\includegraphics[scale=0.3]{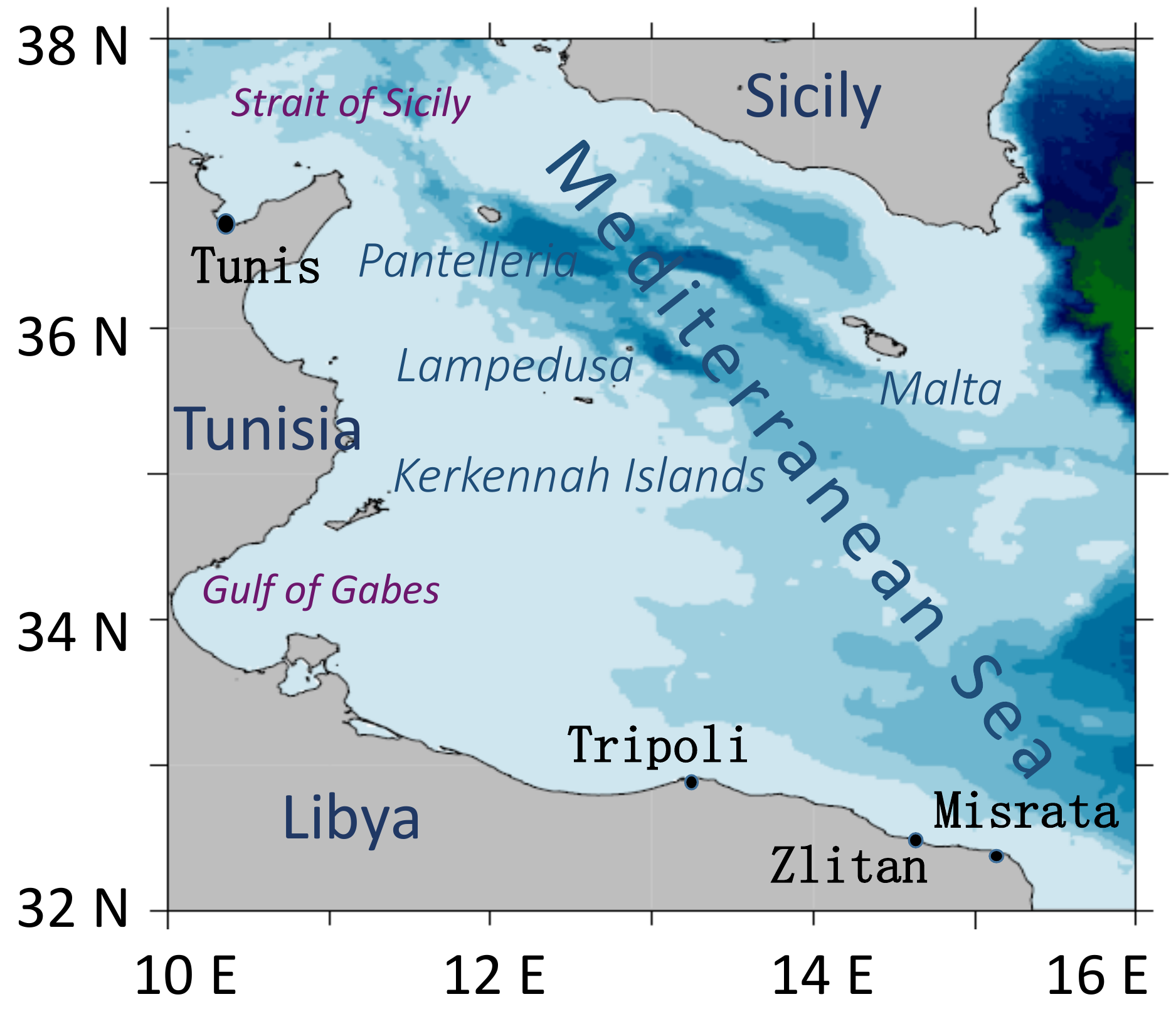}
\caption{Map of the study domain in the CMED.}\label{fig1}
\end{figure}

 Since for the Mediterranean the trajectories of sea drifters largely depend on seasonal ocean surface circulation \citep{Poulain1998} the accumulation of particles is also more likely to be influenced by seasonal and inter--annual ocean and atmospheric wind variability. Here we use short--term simulations to identify seasonal patterns in particles accumulation thus taking into consideration annual and seasonal surface current and wind varibility. We chose a study domain in the CMED that covers area between eastern and western Mediterranean sub--basins (Fig. \ref{fig1}). The CMED extends from the Strait of Sicily and the Malta Channel to the western boundaries of the Ionian Sea, the southern border of the domain includes the coastal areas of Tunis and Libya.     

In our simulations we use Leeway model \citep{Breivik2008,Breivik2012} that is based on the Leeway drift of floating object with respect to ambient current under the influence of wind and waves. In this approach a particle is viewed as a proxy for a surface sea object and the object's trajectory is identified by running a statistical ensemble of particles with uncertainties defined by atmospheric and oceanographic fields and particle characteristics. We use the model that is based on modeling characteristics of sea objects of a lower windage category \citep{AllenPlourde1999} and climatological data from the CMED region to obtain seasonal particle accumulation patterns. The trajectory of a sea object is identified by running a statistical ensemble of particles with uncertainties defined by atmospheric and oceanographic fields as well as particle characteristics. The simulations are performed for $30$ days. This numerical Leeway model \citep{Breivik2008} estimates a final search area using modelling of stochastic Lagrangian trajectories of an ensemble of particles. Identical particles are initially positioned on the regular horizontal lattice covering the study region and are allowed to leave the open simulations domain. The simulations are performed using daily ocean currents from the  Copernicus Marine Environment Monitoring Service (CMEMS)  with horizontal resolution of  $0.0625^\circ \times 0.0625^\circ$ taken at $1.472$ m depth. For the wind forcing we use six--hourly $10$ m atmospheric wind available from the  European Centre for Medium--Range Weather Forecasts (ECMWF) at horizontal resolution of $0.75^\circ \times 0.75^\circ$. We use the mean ocean currents and wind for reference periods in summer and winter of years 2013--2017.

\section{Results}   
\begin{figure}
\includegraphics[scale=0.21]{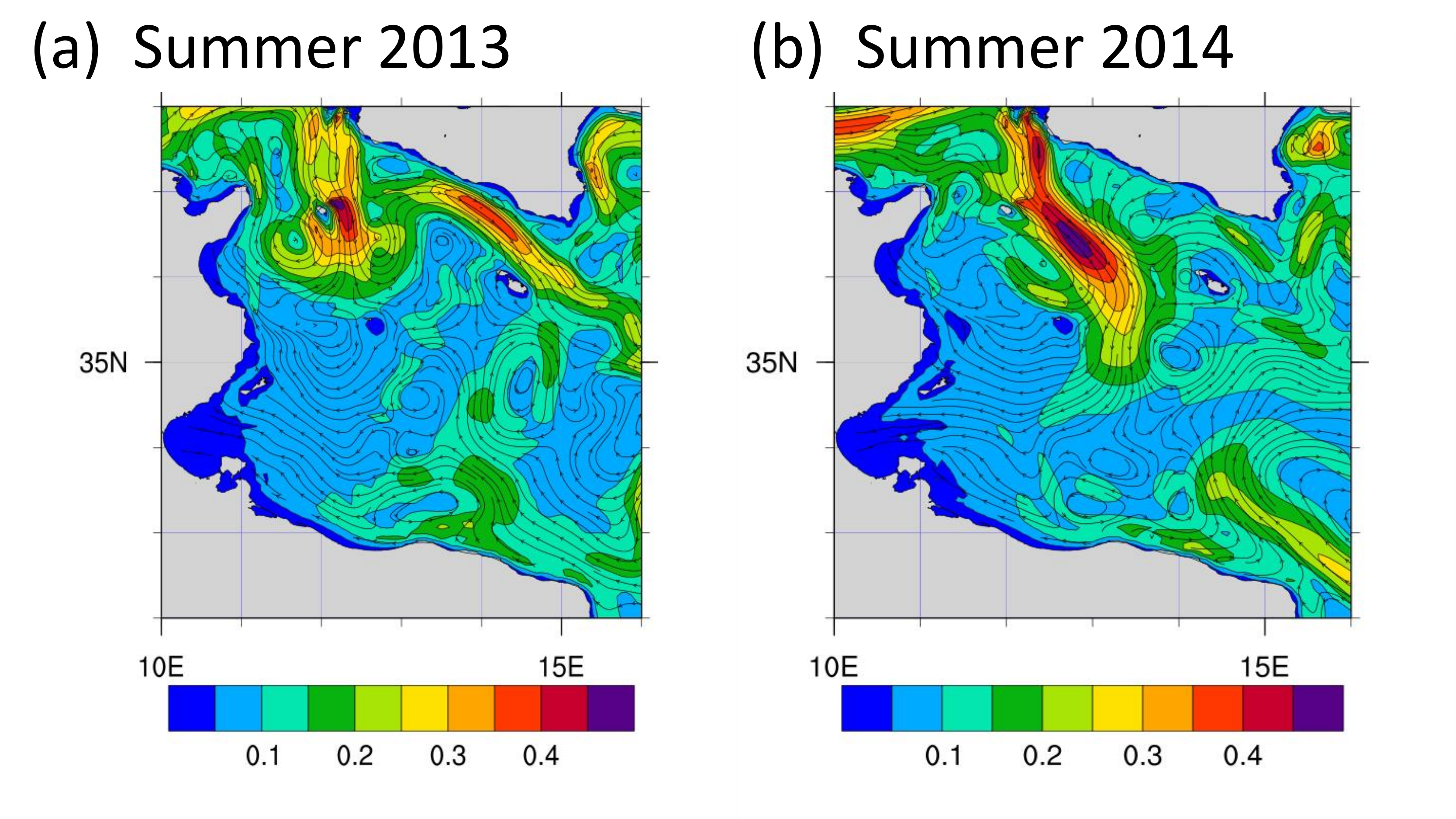}
\includegraphics[scale=0.21]{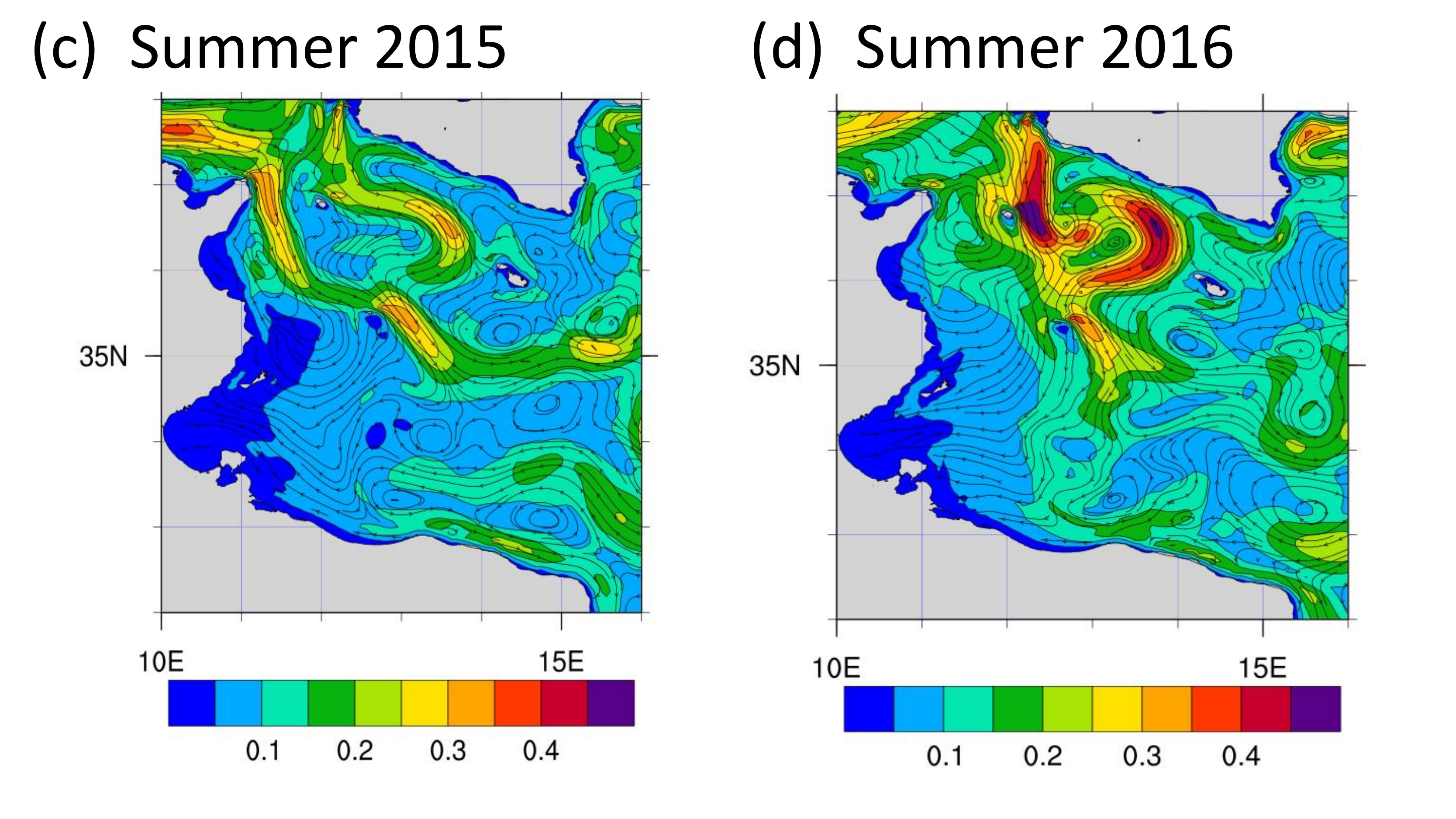}
\includegraphics[scale=0.21]{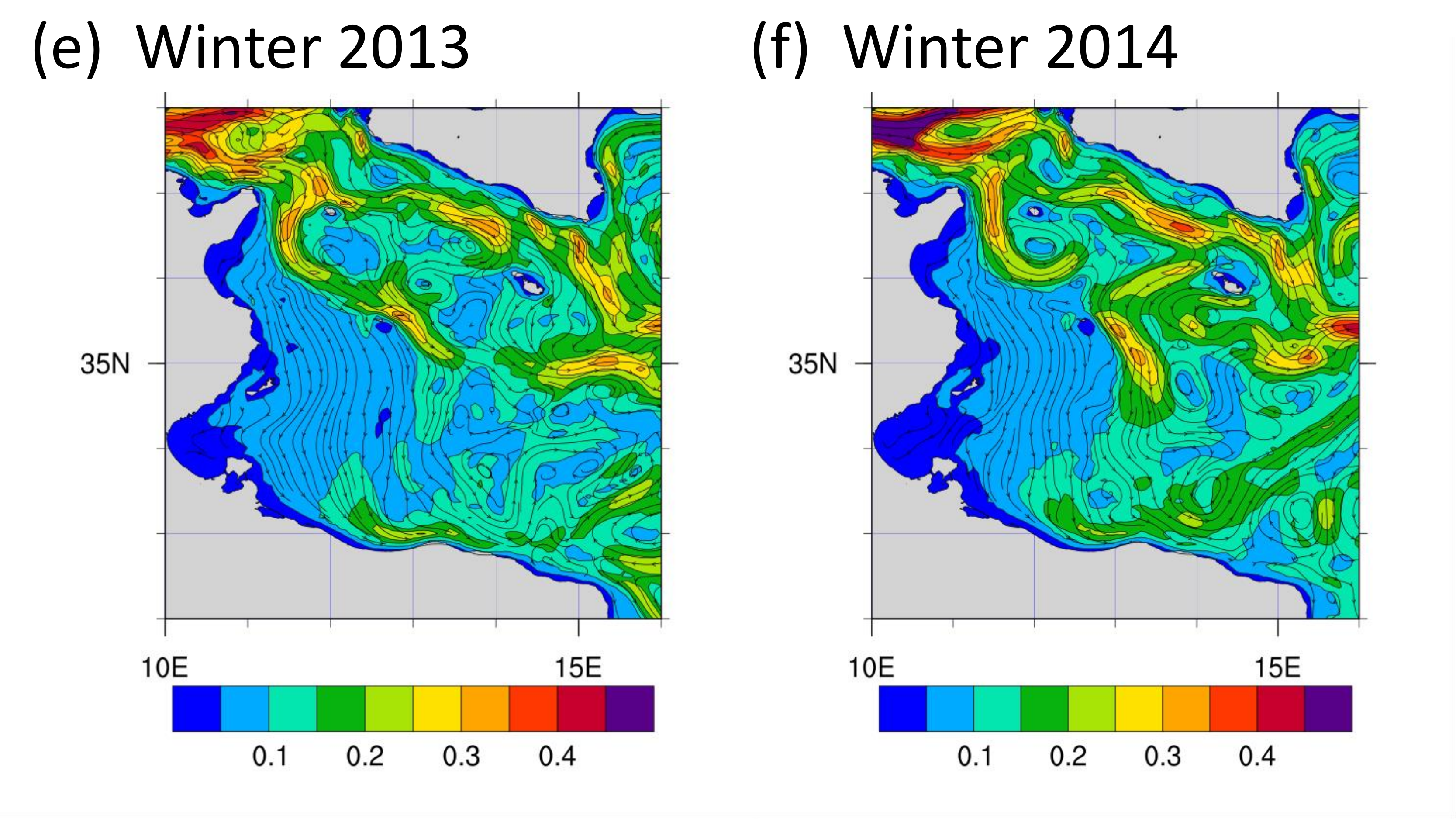}
\includegraphics[scale=0.21]{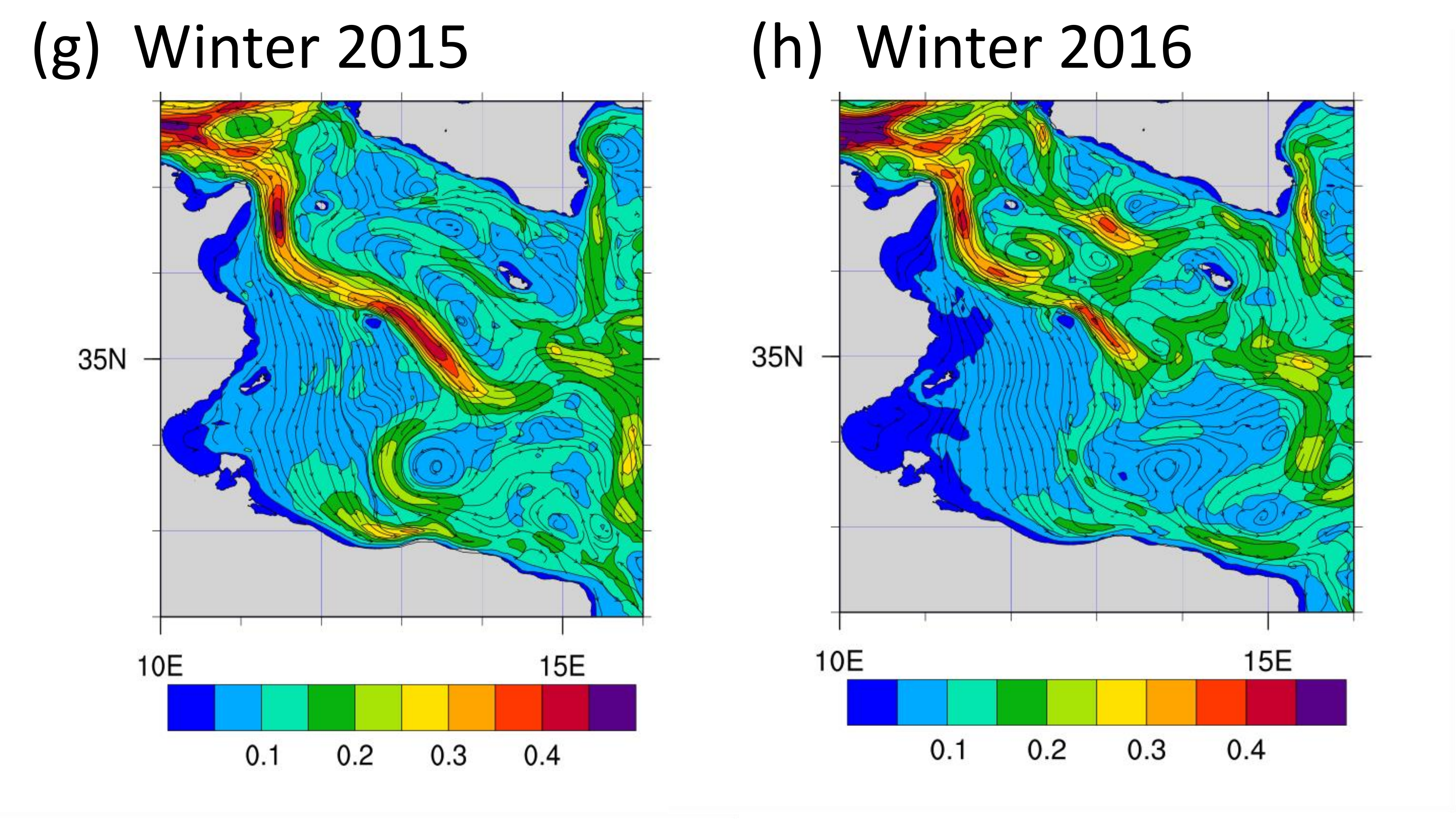}
\caption{Mean MFS surface currents for years 2013--2017. Colorcode and streamlines indicate the three--month means in $[$m/s$]$. Data are retrived from the CMEMS.}\label{fig2}
\end{figure}

The sub--basins circulation in the CMED is dominated by mixing and transformation of a fresh Modified Atlantic Water (MAW) that flows eastward near the surface and a salty deep Levantine Intermediate Water (LIW) that flows in the opposite direction \citep{Pinardi2000,Sorgente2003}. The MAW in the Strait of Sicily is sub--divided into two major branches: the larger branch along the Tunisian coast called Atlantic Tunisian current (ATC) and the smaller branch along the southern Sicilian shelf that further becomes a part of the north Atlantic Ionian Stream (AIS) \citep{Robinson1999}. The AIS was pronounced in years 2013--2014 in winter and in summer (Fig. \ref{fig2} a,e and f). While the ATC as a rule is more pronounced in winter (Fig. \ref{fig2} g and h) during summer 2015 it was stronger than in other years and contributed to the transport of surface water across the domain (Fig. \ref{fig2} c).  In summer 2014 and 2016 the branches of the ATC and AIS bifurcated from its regular circulation and magnified current was observed west of Pantelleria  (Fig. \ref{fig2} b and d).  
\begin{figure}
\includegraphics[scale=0.21]{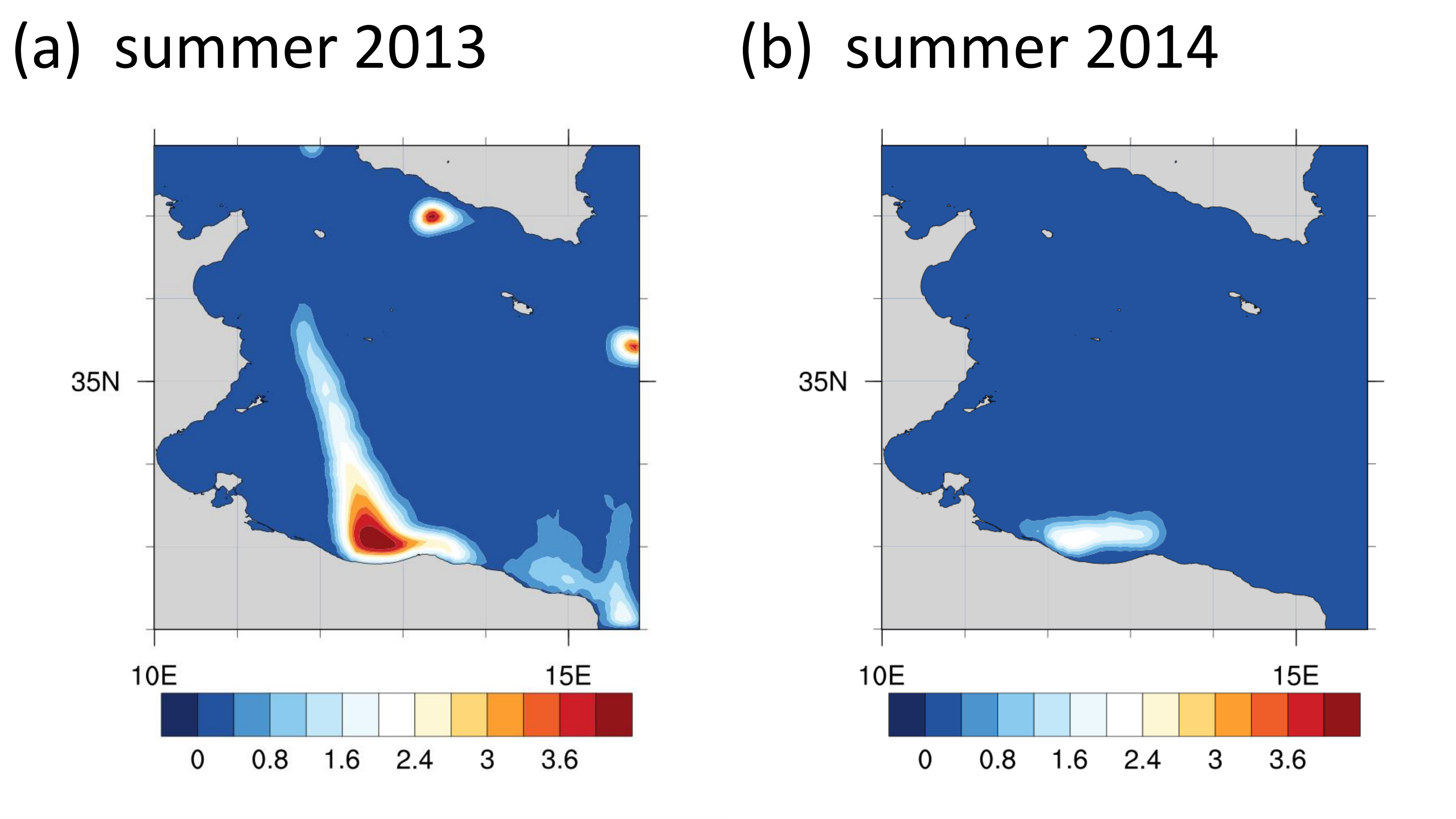}
\includegraphics[scale=0.21]{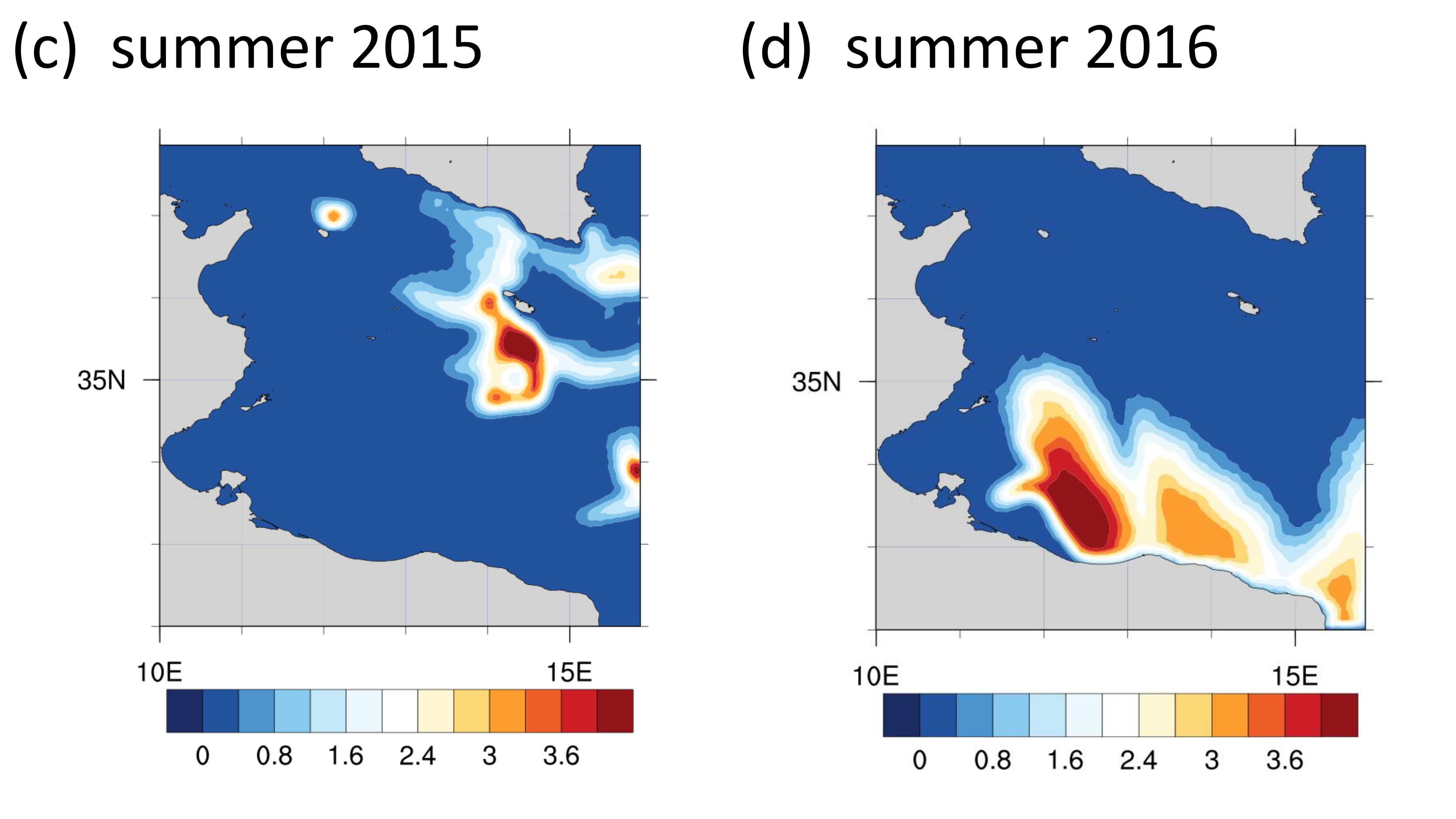}
\includegraphics[scale=0.21]{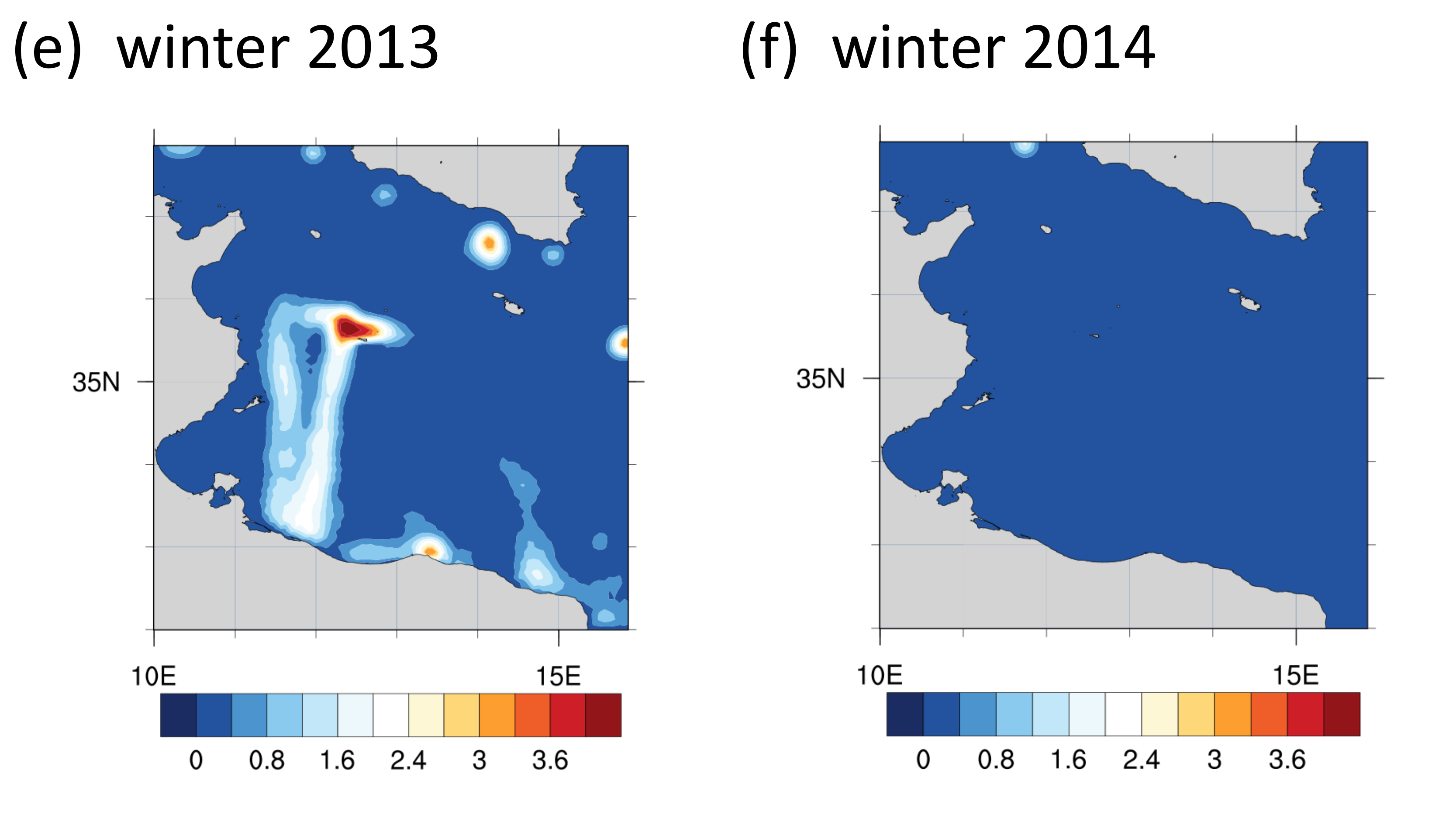}
\includegraphics[scale=0.21]{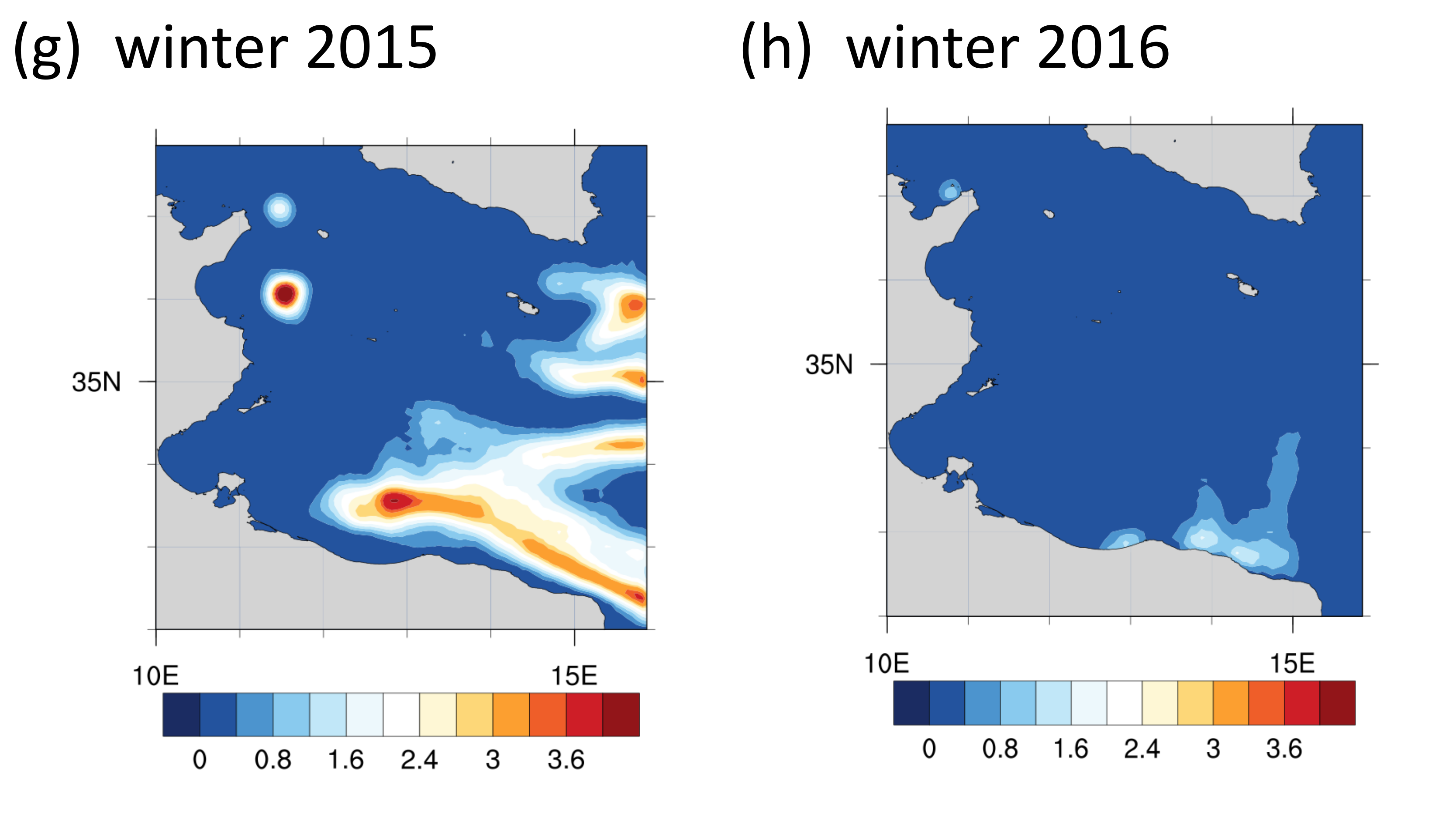}
\caption{Binning surface density for lower windage category. Surface density is  measured after $T=30$ days.  For all experiments $N=16 \times 10^4$ particles are initialized on the regular lattice with the size of a single cell $0.015^\circ \times 0.015^\circ$.  Density of particles is estimated per area of size $\sim36.11$ km$^2$.}\label{fig3}
\end{figure}

Except for the years 2014 and 2015 we found that for summer and winter particles were collected at the north African coast near to major Libyan harbors (Fig. \ref{fig3}). This result is in agreement with the conclusions from Mansui et al. (2014) that found retention area in the proximity of the Libyan coast. The short--term aggregation of particles near to the Libyan coast (Fig. \ref{fig3} a,b,d,e and h) is likely to be caused by the magnification of the ATC and contribution of south easterly wind that carries surface particles to the south of domain. The situation was different during 2014 when intensified AIS transported surface particles across the CMED from the western part of Sicily further to the south (Fig. \ref{fig3} b). While in winter 2014 the ATC was weaker than in later periods the AIS influenced the particle transport along the Sicilian coast towards the Ionian basin quickly, thus, after simulation period most of the particles left the domain (Fig. \ref{fig3} f). The winters of 2015 and 2016 were characterized by a similar ATC pattern, however stronger easterly current at the Libyan coast influenced faster transport of  coastal and near--to--coastal particles towards the Gulf of Sirte and outside of study area(Fig. \ref{fig3} g and h).
\section{Conclusions}
From the monthly simulation of homogeneously distributed particles with characteristics of small size and lower windage drifter we showed that the Libyan coast is in majority of cases impacted by higher concentration of at--sea drifter--like particles. At the same time the positions of hot--spots are affected by annual changes and seasonality of wind and ocean surface currents.     
\section{Acknowledgements}
We are grateful to the CMEMS for the provision of ocean data and to the ECMWF for meteorological data. 
%\section{References}
 \bibliographystyle{elsarticle-num-names} 

\end{document}